\documentclass[aps,pra,twocolumn,amsmath,amssymb,showpacs]{revtex4-1}
\usepackage{graphicx}
\usepackage{amsmath}
\usepackage{bm}

\newcommand{\be}{\begin{equation}}  
\newcommand{\ee}{\end{equation}}
\newcommand{\ba}{\begin{array}}
\newcommand{\ea}{\end{array}}
\newcommand{\bea}{\begin{eqnarray}}
\newcommand{\eea}{\end{eqnarray}}

\newcommand{\bra}{\langle}
\newcommand{\ket}{\rangle}

\newcommand{\nn}{\nonumber}

\begin{document}

\title{Quantum force estimation in arbitrary non-Markovian Gaussian baths} 

\author{C.L. Latune$^1$, I. Sinayskiy$^{1,2}$, F. Petruccione$^{1,2}$}
\affiliation{$^1$Quantum Research Group, School of Chemistry and Physics, University of
KwaZulu-Natal, Durban, KwaZulu-Natal, 4001, South Africa\\
$^2$National Institute for Theoretical Physics (NITheP), KwaZulu-Natal, 4001, South Africa}

\date{\today}
\begin{abstract}
The force estimation problem in quantum metrology with arbitrary non-Markovian Gaussian bath is considered. No assumptions are made on the bath spectrum and coupling strength with the probe. Considering the natural global unitary evolution of both bath and probe and assuming initial global Gaussian states we are able to solve the main issues of any quantum metrological problem: the best achievable precision determined by the quantum Fisher information, the best initial state and the best measurement. Studying the short time behavior and comparing to regular Markovian dynamics we observe an increase of quantum Fisher information. We emphasize that this phenomenon is due to the ability to perform measurements below the correlation time of the bath, activating non-Markovian effects. This brings huge consequences for the sequential preparation-and-measurement scenario as the quantum Fisher information becomes unbounded when the initial probe mean energy goes to infinity, whereas its Markovian counterpart remains bounded by a constant. The long time behavior shows the complexity and potential variety of non-Markovian effects, somewhere between the exponential decay characteristic of Markovian dynamics and the sinusoidal oscillations characteristic of resonant narrow bands.   
\end{abstract}
%\pacs{}

\maketitle

\section{Introduction}
Quantum metrology brought a revolution in parameter estimation theory. The exploitation of subtle quantum resources as entanglement and squeezing offers new ways of dribbling noise without requiring brute force energy increases (often not possible or not welcome). 
An important example is the detection of gravitational wave where quantum metrology allows to reduce the leading remaining noise in the detectors, the almost irreducible quantum fluctuations, by introducing squeezed light (reduction of 28\% of the shot noise without additional noise \cite{natphot}). Note that the recent observation of gravitational waves from a binary Black Hole Merger \cite{ligo} was realized with coherent light, but the next generation experiment should use squeezed light. In such an experiment the strain of the gravitational wave is encoded in a phase difference  between the electromagnetic fields running in the two arms of an interferometer. The resulting detection task is an estimation of rotation in phase space.

 Here, we focus on force estimation, which corresponds to estimation of displacement in phase space. High precision force estimation can be implemented in several setups like optomechanics \cite{impulsiveforce, microopto, aspel}, trapped ions \cite{biercuck}, atomic force microscopy \cite{ieee} and others harmonic oscillators \cite{nanotube}, useful for testing or exploring fundamental physical phenomena \cite{singlespin, pt, ca, planck,nori}, and in Biology \cite{biology}.

 The treatment of force estimation problem with realistic environment including non-Markovian noise have been pioneered in \cite{euroj}. Most of the publications dealing with non-Markovian noise concern two-level system quantum metrology \cite{plenio,matsuzaki,dd}. 

In this paper we treat the force estimation in presence of an environment constituted of a collection of harmonic oscillators without any assumption on the bath spectrum and assuming linear coupling of any strength with the probe. We develop a new approach to study the non-Markovian effects of the bath. Ideally the dynamical parameter to be estimated is encoded in the system or $probe$ via unitary evolution. For a realistic process noise should be added. Rather than constructing and solving a dissipative master equation obtained eventually from several approximations on the probe-environment coupling and initial state, we consider the global unitary evolution of the system and its environment. 
In doing so we avoid the traditional Markovian and weak coupling approximations and the resolution of a master equation. We start directly from a physical purification of the probe dynamics determined by the bath properties and bath-probe coupling coefficients. We also avoid problems related to potential initial correlations between probe and bath which can yield non-completely positive maps when tracing out the bath \cite{pechukas}.

In this work, effects which arise exclusively from the dynamics at times smaller than the bath correlation time are called non-Markovian. In other words, being able to make measurement at time scales below the bath correlation time can potentially unveil new effects that we call non-Markovian. This notion of non-Markovianity meets its classical counter-part since it refers to the memory of the bath defined as the correlation time. 
 How this is related to more elaborated definitions which intend to characterize non-Markovianity by short, medium or long time effects is a highly non-trivial question. Hereafter we refer to non-Markovian noise as noise appearing below the correlation time of the bath.

The parameter estimation protocol is the following. The probe system is a harmonic oscillator $S$. After a time window $[t_0;t]$ of force sensing (interaction between the probe and the force), the probe is measured and the force is estimated from the measurement output given that the initial state is known. The function which generates an estimate from the measurement outputs is called estimator. To obtain a better estimate the process has to be repeated several times and the estimate updated via the chosen estimator. For a given measurement the maximal achievable precision of the estimation is determined by the so-called Fisher information \cite{fisher,cramer,rao}. We denote by $F$ the force intensity, $\rho_S(t,t_0,F)$ the probe state just before the measurement, $\{\Pi_m\}_m$ a POVM (positive-operator-valued-measure) describing the measurement, and $p(m|F) = {\rm Tr}[\rho_S(t,t_0,F)\Pi_m]$ the conditional probability of getting the measurement output $m$. The Fisher information associated to this estimation experiment is \cite{fisher}
\be
{\cal F}(\{\Pi_m\}_m,F) = \int dm \frac{1}{p(m|F)}\left[\frac{\partial p(m|F)}{\partial F}\right]^2.
\ee
The Fisher information as well as the precision of the force estimation depend strongly on the initial state of the probe and on the measurement applied. The maximal Fisher information over all possible POVM is called the quantum Fisher information (QFI) ${\cal F}_Q(F) = {\rm Max}_{\{\Pi_m\}_m}\{{\cal F}(\{\Pi_m\}_m,F)\}$ . The uncertainty of the estimation is characterised by the mean square of the error $\delta F$ between the estimate $F_{est}$ and the real value $F$ and is lowered by the Cramer-Rao bound \cite{cramer,rao}:
\be\label{qcrb}
\langle\delta^2 F\rangle:= \langle(F_{est}-F)^2\rangle \geq [\nu {\cal F}_Q(F)]^{-1},
\ee
where $\nu$ is the number of independent repetitions of the experiment and $\langle...\rangle$ is an ensemble average. This lower bound is saturated for efficient and unbiased estimators \cite{fisher,caves,bcm96} and when best measurements are realised. Identifying best measurements and efficient estimators is in general a complex task. Except for special situation, there is no efficient estimator for finite $\nu$ \cite{bcm96}. The inequality \eqref{qcrb} is sometimes referred as local estimation not only because at principle the lower bound depends on the value of the parameter $F$ and consequently characterises the precision only at this value but also because estimators which saturate the bound may depend on the value of the unknown parameter and their efficiency may require that the range of the possible values for the unknown parameter is previously known. 
However, because we choose a linear interaction between the force and the probe, the QFI does not depend on $F$ and the lower bound of the error function (see \cite{tsang}) would remain the same as the actual Cramer-Rao bound. Note also that the hypothesis of initialising the probe and bath in a global Gaussian state guarantees that the distributions generated by quadrature measurements are Gaussian, and then the maximum likelihood estimator \cite{fisher,braunstein} is efficient from the first measurement ($\nu=1$).
 The QFI determines the maximal achievable precision turning it the central quantity in most works in quantum metrology.\\  

In the following we derive an analytic expression of the QFI and determine the best initial state and the best measurement with the only assumption of Gaussianity of the initial global state of the probe and environment. From this result we show that surprisingly, at short times $t-t_0$ below the bath correlation time, the QFI is equal up to order 3 in $t-t_0$ to the QFI for noiseless evolution, whereas the QFI with Markovian approximation is equal to the noiseless expression only up to order 2 in $t-t_0$. This interesting feature is responsible for a huge difference between Markovian and non-Markovian noise regimes for what we call sequential preparation-and-measurement scenario, a repetition of sequence of probe state preparation, sensing of the force for an adequate duration $\tau$, and  measurement (Section \ref{seqmes}). We show that with growing initial energy of the optimally
prepared probe, the optimal duration of the protocol-step diminishes,
so that by entering timescales below the bath correlation time one may
benefit from the non-Markovian effects. As a result, the QFI goes to
infinity (and the uncertainty of the estimation to zero) whereas it remains limited by a constant for protocol-step timescales bigger than the bath correlation time, where the probe experiences  Markovian noise. \\
In \cite{dd} the authors observe that the QFI for frequency estimation by qubits can diverge with growing initial energy. Their observation points in the same direction as ours, but they conclude that this super-classical precision is not related to non-Markovianity since they don't observe back-flow of information from the bath to the probe. 
Here we show that the same super-classical precision is attainable in continuous variable systems, and without specific hypothesis on the noise as in \cite{dd}, just assuming Gaussianity of the bath (which includes ohmic, sub-ohmic or super-ohmic, a "general environment" as called in \cite{paz}).
Although back-flow of information is a witness of non-Markovianity, it is not a universal criteria. In \cite{dd} it is not possible to analyse directly the bath memory properties because the bath does not appear explicitly. In our model we can directly compare the bath dynamics timescales with the measurement timescale and conclude that the super-classical precision occurs when the measurement timescale enters the memory bath timescale, where non-Markovian noise arises.

\section{Global dynamics}\label{global}
As mentioned in the introduction the harmonic oscillator $S$ is linearly coupled to the force to be estimated. The force is modulated by a time function $\zeta(t)$ assumed to be known. The global unitary evolution of the probe $S$ and the bath $B$ is denoted by $U(t,t_0,F)$. 
We use hereafter the following notations: 
\bea
X(\theta)&:=&\frac{1}{\sqrt{2}}(a^{\dag}e^{i\theta} + a e^{-i\theta}),\nn\\ 
X(\theta) [t,t_0,F]&:=&U^{\dag}(t,t_0,F)X(\theta)U(t,t_0,F),\nn\\
\big\bra X(\theta)[t,t_0,F]\big\ket_0&:=&{\rm Tr}_{SB}\{\rho_{SB}^0  X(\theta)[t,t_0]\},\nn\\
\big\bra\Delta^2 X(\theta)[t,t_0]\big\ket_0&:=& \big\bra \{X(\theta)[t,t_0]\}^2\big\ket_0 -\big\bra X(\theta)[t,t_0]\big\ket_0^2.\nn
\eea
 Note that defined in this way $X(\theta)[t,t_0]$ is an observable of both probe and bath's Hilbert space. By taking the expectation value in the state $\rho_{SB}^0$ one recovers the usual expression for the expectation value of a probe observable. Note also that the quantity inside the parenthesis (here $\theta$) and $[t,t_0]$ are independent: the first one designates the quadrature angle, which in the following will be chosen according to the instant of measurement $t$, and the second one designates the time evolution of this same quadrature between $t_0$ and $t$. 

The global Hamiltonian is given by:
\begin{eqnarray}
H(t,F)/\hbar &=& \omega_0 a^{\dag} a - F\frac{\omega_0}{\sqrt{2}}\zeta(t)\left(a^{\dag}+a\right) + \sum_n \omega_n b_n^{\dag}b_n \nonumber\\
&&- ia^{\dag}\sum_n K_n b_n +ia\sum_n K_n^{*}b_n^{\dag} ,
\end{eqnarray}
where $F$ is the force intensity, the parameter to estimate. We consider that the environment is constituted of a collection of harmonic oscillators coupled to $S$ via the coefficient $K_n$. We show (Appendix \ref{evolution}) that the global unitary interaction $U(t,t_0)$ can be written in the following form
\bea\label{evolop}
U(t,t_0,F) = &e^{-\frac{i}{\hbar}(t-t_0)H_0}&e^{iF|D(t,t_0)|X(\phi^D_{t,t_0})} R(t,t_0){\cal B}(t,t_0),\nn\\
\eea
where $t_0$ designates the instant of time of the beginning of the interaction of the probe with the bath and the force. Expression \eqref{evolop} shows that we can split $U(t,t_0,F)$ in a succession of four operations  (see details in Appendix \ref{evolution}). Firstly a probe-bath mixing ${\cal B}(t,t_0)$. Then a displacement proportional to $F$ implemented by the operator $R(t,t_0)$ in the phase space of the bath. A second displacement given by $e^{iF|D(t,t_0)|X(\phi^D_{t,t_0})}$ and taking place in the phase space of $S$,  with  $D(t,t_0) = \omega_0\int_{t_0}^tdu \zeta(u) e^{i\omega_0(u-t_0)}G(t,u) $, $\phi^D_{t,t_0}=\arg{D(t,t_0)}$, and the complex function $G(t,u)$ describing the bath response. Its explicit expression is similar to a Dyson series and is shown at the beginning of Appendix \ref{evolution}. The last operation is the free evolution $e^{-\frac{i}{\hbar}(t-t_0)H_0}$ with $H_0/\hbar=\omega_0 a^{\dag} a+ \sum_n \omega_n b_n^{\dag}b_n$.

Expression \eqref{evolop} is by itself an important and interesting result by its simple form. It allows exact derivation of probe and bath observables in Heisenberg picture and corresponding moments (Appendix \ref{evolution}). Usual treatment would require much more work to derive those quantities: establish an exact non-Markovian master equation and then solve it. With our method we have also the possibility of deriving those quantities without hypothesis of initially uncorrelated bath and probe, which would be hardly possible by a master equation treatment \cite{pechukas}. However this possibility is not interesting in the present work since we focus on the best initial state which is a pure state.

Note that the expression \eqref{evolop} is similar - without $R(t,t_0)$ - to the form found in \cite{pra} where the noise is treated via a Markovian master equation. Note also that \eqref{evolop} is obtained without doing any approximation regarding the strength of the coupling with the bath, and the bath spectral density.

\section{Derivation of the metrological quantities for Gaussian states}
In order to access the QFI of $S$ related to the parameter $F$ we have to maximize the Fisher information ${\cal F}(\{\Pi_m\}_m,F)$ over all physical measurements $\{\Pi_m\}_m$. Unfortunately this is not tractable. Efforts have been made to find alternatives \cite{caves, paris, bruno, nicim}, but they are still hard to apply for arbitrary dynamics and states. We  restrict ourselves to Gaussian states which are more easily accessible experimentally and for which \cite{pinel, monras, jiang} have derived explicit expressions of the QFI. As we will see in the following, this important assumption also guarantees that the best measurement is a quadrature measurement and that the maximum likelihood estimator is efficient implying that the quantum Cramer-Rao bound Eq. \eqref{qcrb} is saturated for any number of repetitions of the experiment ($\nu \geq 1$) \cite{fisher,braunstein}. Assuming that the initial global state of the probe and bath $\rho_{SB}^0$ is Gaussian implies that at any time the probe state $\rho_S(t,t_0)$ is Gaussian too since the global Hamiltonian $H(t,F)$ is quadratic and the partial trace is a Gaussian operation. We are now able to derive an expression of the QFI thanks to results from \cite{caves} and expressions for fidelity between Gaussian states \cite{scutaru}. One can also use the derivations in \cite{pinel, monras, jiang}  which are applications of the results in \cite{caves} to Gaussian states. As expected, all these expressions of the QFI for Gaussian states are functions of the first and second moment of the quadrature operators. 
As already mentioned at the end of Section \ref{global} these quantities are highly non-trivial to derive without approximations. Here we do so  (Appendix \ref{evolution}) thanks to the simple form of \eqref{evolop}. 
Since in this problem of force estimation the parameter $F$ to be estimated controls only the amplitude of the displacement in the phase space, the expression for QFI derived in \cite{pinel} for Gaussian states can be simplified to
\bea\label{firstqfi}
{\cal F}_Q (t,t_0) &=& \frac{ |D(t,t_0)|^2}{{\rm Det}{\bf\Sigma}(t,t_0)}\big\langle\Delta^2 X(\phi^D_{t,t_0} - \omega_0(t-t_0))[t,t_0]\big\rangle_0,\nn\\
\eea
where ${\rm Det}{\bf \Sigma}(t,t_0)$ is the determinant of the covariance matrix ${\bf \Sigma}(t,t_0)$ of $\rho_S(t,t_0)$. Note that the use of the result of \cite{pinel} requires that the initial probe state is Gaussian, but this does not exclude correlation with the bath. The minimum condition is indeed that $\rho_{SB}^0$ be Gaussian. For $\rho_S(t,t_0)$ being Gaussian, the determinant ${\rm Det}{\bf \Sigma}(t,t_0)$ is equal to the product of the extremal quadrature variances. Let's call $\theta_m^t$ the angle of the maximal quadrature variance at time $t$ so that we have ${\rm Det}{\bf \Sigma}(t,t_0) = \big\langle\Delta^2 X(\theta_m^t)[t,t_0]\big\rangle_0 \times\big\langle\Delta^2 X(\theta_m^t+\pi/2)[t,t_0]\big\rangle_0$. The angle $\theta_m^t$ is a function of $t$ and of $\theta_m^0$, the angle of the maximal quadrature variance of the initial state at $t_0$. If the value of $t$ is predetermined, meaning that we choose the length of the sensing window before the beginning of the experiment (as for sequential preparation-and-measurement scenario, Section \ref{seqmes}), $\theta_m^t$ is uniquely determined by $\theta_m^0$ and so there exists one $\theta_m^0 \in[0,\pi[$ such that $\theta_m^t = \phi^D_{t,t_0} - \omega_0(t-t_0)$. As shown in Appendix \ref{evolution}, for the probe and bath initially uncorrelated, the simple following relation holds $\theta_m^t = \theta_m^0 + \phi_{t,t_0}^G -\omega_0(t-t_0)$, where $\phi_{t,t_0}^G:=\arg{G(t,t_0)}$. Under these conditions, preparing the probe with $\theta_m^0 = \phi_{t,t_0}^D-\phi_{t,t_0}^G$ guarantees that $\theta_m^t = \phi^D_{t,t_0} - \omega_0(t-t_0)$, and consequently the expression of the QFI reduces to
\be\label{qfipvar}
{\cal F}_Q(t,t_0) = \frac{ |D(t,t_0)|^2}{\Big\langle\Delta^2 P(\phi^D_{t,t_0} - \omega_0(t-t_0))[t,t_0]\Big\rangle_0}.
\ee
This expression is remarkable by its simplicity and similarity to the noiseless and Markovian dynamics expressions \cite{pra}. Note that the condition $\theta_m^0 = \phi_{t,t_0}^D-\phi_{t,t_0}^G$ is obviously satisfied for any circularly symmetric states like coherent or Fock states. 

However, more relevant than being able to write the QFI in a simple form is the best initial state. We already know that it belongs to the pure states (because of the convexity of the QFI), which by the way implies that the best state is not correlated to the bath. Maximizing \eqref{firstqfi} implies maximizing the variance $\big\langle\Delta^2 X(\phi^D_{t,t_0} - \omega_0(t-t_0))[t,t_0]\big\rangle_0$ (given a limited initial mean energy $E$) and minimizing ${\rm Det}{\bf \Sigma}(t,t_0)$ which is already minimal and equals to $1/4$ if $S$ is initialized in a pure state. In Appendix \ref{evolution} we show that the best state is the squeezed state $\hat S[\mu(t,t_0)]|0\rangle$ where $|0\rangle$ is the ground state of the harmonic oscillator, $\hat S[\mu(t,t_0)] = \exp{\left(\frac{\mu(t,t_0)}{2} a^{\dag 2} - \frac{\mu^{*}(t,t_0)}{2} a^{2}\right)}$ is a squeezing operator, with $\mu(t,t_0)=re^{2i(\phi^D_{t,t_0}-\phi^G_{t,t_0})}$ and $r=\frac{1}{2}\ln{[2(E+\sqrt{E^2-1/4})]}$. One can see that the best state is squeezed along the quadrature $P(\phi_{t,t_0}^D-\phi_{t,t_0}^G)$ implying that the condition $\theta_m^0 = \phi_{t,t_0}^D-\phi_{t,t_0}^G$ is satisfied and that the form of the QFI for the best state is also \eqref{qfipvar}. Substituting the variance of the quadrature $P(\phi^D_{t,t_0} - \omega_0(t-t_0))[t,t_0,F]$ by its expression for the best state gives
\begin{widetext} 
\bea\label{qfibeststate}
{\cal F}_Q(t,t_0) = \frac{ |D(t,t_0)|^2}{|G(t,t_0)|^2\left[4(E+\sqrt{E^2-1/4})\right]^{-1}+\sum_n |K_n|^2\left(N_n + \frac{1}{2}\right) \Big|\int_{t_0}^tds G(t,s) e^{i(\omega_0-\omega_n)s}\Big|^2},
\eea
\end{widetext}
where $N_n := {\rm Tr}[\rho_B^0b_n^{\dag}b_n]$.
This expression shows the transition between noiseless ($K_n=0$ and $G(t,t_0)=1$) and noisy quantum metrology. Without noise we recover the so-called Heisenberg limit characterized by a linear dependence in $E$, leading to an infinite precision for growing $E$ as in noiseless classical parameter estimations. For noisy quantum metrology, as time goes the influence of the bath grows ($|G(t,t_0)| \leq 1$) and can eventually dominate the initial preparation conditions ($|G(t,t_0)| \ll 1$) spoiling efforts to recover infinite precision. For broadband limit and rotating wave approximation $G(t,t_0)$ tends to 0 at long times (see Appendix \ref{bnband}), erasing the initial conditions dependence and recovering the Markovian behavior \cite{pra}. We also show in Appendix \ref{bnband} that the resonant narrow band limit reduces $G(t,t_0)$ to a cosine resulting in an indefinite succession of forward and backward flows of information between $S$ and the bath along with an indefinite dependence on the initial conditions, contrasting with the Markovian behavior. However, as we show in the following, the long time behavior of $G(t,t_0)$ cannot be determined in general without explicit expressions of $K_n$, but a behavior between these two extremes of total loss and total recovering of initial information is expected. \\

To complete the protocol of best estimation we determine in Appendix \ref{quadmeas} that the projective measurement onto the quadrature $P(\phi^D_{t,t_0} -\omega_0(t-t_0))$ yields a {\it Fisher information} equal to the right hand side of \eqref{qfipvar}. One concludes that whenever the {\it QFI} is given by \eqref{qfipvar}, in other words whenever the initial state is prepared in a Gaussian state with the maximal variance of the quadrature occurring for $\theta_m^0  = \phi_{t,t_0}^D-\phi_{t,t_0}^G$ (this includes the best state), the best measurement is the projection onto the quadrature $P(\phi^D_{t,t_0} -\omega_0(t-t_0))$. This useful result shows also that the best measurement generates Gaussian distributions which implies that the maximum likelihood estimator is efficient for arbitrary number of repetitions of the experiment \cite{fisher,braunstein}. One can show that in this problem the maximum likelihood estimator is the simple average of the outcomes (successively obtained after each repetition of the experiment) of the projective measurement just described above.

Interestingly we can treat the situation where the instant of beginning and end of the force are not known exactly. Let's call $t_i$ and $t_f$ such instants so that $\zeta(t)=0$ for $t\leq t_i$ and $t\geq t_f$. The only change in the previous expressions is $D(t,t_0) =  \omega_0\int_{t_i}^{t_f}du \zeta(u) e^{i\omega_0(u-t_0)}G(t,u) = D(t_f,t_i)$. As expected, initializing the experiment before the force begins and measuring after the force stops is prejudicial for the precision of the estimation since  $\big\bra \Delta^2 P(\phi^D_{t_f,t_i} -\omega_0(t_f-t_i))[t_f,t_i]\big\ket \leq \big\bra \Delta^2 P(\phi^D_{t_f,t_i} -\omega_0(t-t_0))[t,t_0]\big\ket$. This is one of the arguments for opting for sequential preparation-and-measurement scenario which, as shown in Appendix \ref{seqmeasurement}, avoids the potential problem of knowing $t_i$ and $t_f$.\\

\section{Short and long time behavior}\label{shortandlong}
We assume that we can perform a measurement at a time $t$ such that $t-t_0$ is much smaller than $\omega_0^{-1}$, $\Omega_p^{-1}$, and $|\chi_q|^{-1}$, for all $p\geq2$ and $q\geq1$ (see Appendix \ref{considtimescale}). Under this condition one can legitimately expand Eq.\eqref{qfipvar} around $t_0$:
\bea
{\cal F}_Q(t,t_0) &= \frac{\omega_0^2(t-t_0)^2}{\left\bra \Delta^2 P\left[\phi_{t,t_0}^{D_0}\right]\right\ket_0 }&\left[\zeta^2(t_0) +\zeta(t_0)\dot{\zeta}(t_0)(t-t_0)\right]\nn\\
 &&+ {\cal O}[\omega_0^2{\cal K}^2(t-t_0)^4],\label{fexpanded}
\eea
 where $\phi^{D_0}_{t,t_0}:=\arg{D_0(t,t_0)}$ with $D_0(t,t_0):=\lim_{|K_n|\rightarrow0} D(t,t_0)$, the noiseless value of $D(t,t_0)$ (Appendix \ref{considtimescale}). Consequently the bath dependence appears only at the 4th order in $(t-t_0)$ through the coefficient ${\cal K}^2:=\sum_n |K_n|^2$. One can show from \eqref{firstqfi} that this property is not merely an artefact of some particular initial states but in fact remains valid for any initial Gaussian state. This means that the QFI is equal up to order 3 in $(t-t_0)$ to the QFI without any noise. This is a surprising and interesting result reminiscent of quantum Zeno effect. We show in Appendix \ref{bathcorrelation} that the evolution time scales of the bath correlation function is of order of $\Omega_p^{-1}$ and $|\chi_q|^{-1}$, for all $p\geq2$ and $q\geq1$. This implies that the expansion \eqref{fexpanded} is valid if the measurement is performed at time scales below the correlation time of the bath. \\
If at contrary we cannot perform measurements below the correlation time of the bath, the expansion \eqref{fexpanded} is not valid (the first terms are not significant anymore and the higher terms have to be taken into account). We show in Appendix \ref{shorttime} that in such a situation the short time expansion contains a bath contribution at the 3rd order. The same phenomenon happens for Markovian dynamics.\\

 We show in Section \ref{seqmes} that this short time behaviour is responsible for a great increase (and even a change of scaling) of QFI in sequential preparation-and-measurement scenario protocols when the optimal time interval between each measurement becomes smaller than the bath correlation time scale. \\

 The long time analysis is much more complicated in general due to $G(t,t_0)$. Beginning to look at the long time behavior of the first term of $G(t,t_0)$, one finds that the real part tends to its Markovian equivalent $-\gamma(t-t_0)/2$ and the imaginary part is more involved, but cancels out if $g(\omega)|K(\omega)|^2$ is symmetrical with respect to $\omega_0$ (see Appendix \ref{longtime}).
However, one can show that even in the limit of $t-t_0$ much bigger than all time scales involved in the problem the second term in the expansion of $G(t,t_0)$ is different form its Markovian equivalent, $\gamma^2(t-t_0)^2/8$. So the Markovian behavior does not seem to be recovered in the long time limit as it is sometimes claimed \cite{euroj}. Note that for arbitrary $g(\omega)|K(\omega)|^2$ the imaginary part does not cancel out and its contribution can give rise to diverse long time behaviors far different from the Markovian one as for instance the extreme case of resonant narrow band (Appendix \ref{bnband}).

\section{sequential preparation-and-measurement scenario}\label{seqmes}
 Let $[t_0,t_0 + T]$ be the supposed time window available for the sensing of the force. In order to increase the Fisher information about $F$ one can repeat the measurement $\nu$ times after a sensing time interval of $\tau$ such that $\nu \tau =T$. This process increases significantly the precision of the estimation when the window duration $T$ is bigger than the relaxation time of the probe. The time interval $\tau$ must be chosen adequately. There are two competing quantities growing with time: the information about $F$ in the probe state and the environment-induced fluctuation of the probe state. We are interested now in determining this optimal time interval $\tau_{\mathrm{opt}}$.\\

The total QFI is just the sum of each Fisher information generated after each measurement at time $t_k:=t_0+k\tau$:
\bea\label{fseq}
{\cal F}_Q^{\mathrm{Seq}}(T,\tau) &=& \sum_{k=0}^{\nu-1} {\cal F}_Q(t_{k+1},t_k)\nn\\
&=& \sum_{k=0}^{\nu -1}\frac{|D(t_{k+1},t_k)|^2}{\bra \Delta^2 P(\phi^D_{t_{k+1},t_k} -\omega_0\tau)[t_{k+1},t_k]\ket_0}.
\eea

The second line is valid if after each measurement the probe is prepared in the best state $\hat S[\mu(t_{k+1},t_k)]|0\rangle$. This is what we call sequential preparation-and-measurement scenario. Once we choose $\tau$ we can  evaluate (at least numerically) $\phi^D_{t_{k+1},t_k}$ and prepare the state $\hat S[\mu(t_{k+1},t_k)]|0\rangle$. The challenge is to prepare it in a time interval much smaller than the time evolution of the harmonic oscillator so that this time interval needed for the state preparation can be neglected. With this assumption the variances $\bra \Delta^2 P(\phi^D_{t_{k+1},t_k} -\omega_0\tau)[t_{k+1},t_k,F]\ket$ become k-independent (see Appendix \ref{seqmeasurement}) and can be taken out of the sum in Eq. \eqref{fseq}.\\

To continue the analytic analysis we assume that we are looking for small optimal time interval $\tau_{\mathrm{opt}}$. Assuming that in Eq. \eqref{fseq} $\tau$ is much smaller than $\omega_0^{-1}$, $\Omega_p^{-1}$, and $|\chi_q|^{-1}$, for all $p\geq2$ and $q\geq1$ (see Appendix \ref{considtimescale}), we can legitimately derive a small time expansion (shown in Appendix \ref{seqmeasurement}). From it we deduce,
 \bea\label{optimaltime}
\tau_{\mathrm{opt}}=\frac{1}{2\sqrt{3{\cal N}}}{\cal E}^{-1/2} +  {\cal O}({\cal E}^{-3/2}),
\eea
with ${\cal E}:=\left(E+\sqrt{E^2-\frac{1}{4}}\right)$ and ${\cal N} := \sum_n|K_n|^2(N_n+1/2)$. The second order term is detailed in Appendix \ref{seqmeasurement}. This result illustrates the close relation between optimal time and initial energy $E$: when $E$ goes to infinity, the initial fluctuation of the probe state goes to zero (for the chosen quadrature) and $\tau_{\mathrm{opt}}$ tends to zero. One can derive a sufficient condition on the initial energy $E$ which guarantees that $\tau_{\rm opt} \leq \omega_0^{-1}$, $\Omega_p^{-1}$, and $|\chi_q|^{-1}$, for all $p\geq2$ and $q\geq1$
(see Appendix \ref{considtimescale}). Interesting for experimental implementation, the leading term of Eq. \eqref{optimaltime} does not depend on the total available sensing window $[t_0;t_0+T]$ neither on the force time modulation $\zeta(t)$. For Markovian bath the dependence of $\tau_{\mathrm{opt}}$ is proportional to ${\cal E}^{-1/3}$ \cite{pra}.\\

The corresponding total QFI is
\bea
&&{\cal F}_Q^{\mathrm{Seq}}(T,\tau_{opt})= \frac{\sqrt{3}\xi(T,t_0)}{2\sqrt{\cal N}}{\cal E}^{1/2} + {\cal O}({\cal E}^{-1/2}),
\eea
with $\xi(T,t_0):=\int_{t_0}^{t_0+T}dt\zeta^2(t)$. The second order term is also detailed in Appendix \ref{seqmeasurement}.
The total quantum  Fisher information scales as $E^{-1/2}$ and as $E_{\mathrm{tot}}^{1/3}$ for the total energy  $E_{\mathrm{tot}}:=\nu_{\mathrm{opt}}E = TE/\tau_{\mathrm{opt}}$. This is valid when the initial energy invested in the squeezing of the probe is sufficient (see Eq. \eqref{optimaltime}) so that the optimal time become smaller than $\Omega_p^{-1}$, and $|\chi_q|^{-1}$, for all $p\geq2$ and $q\geq1$, which correspond to the bath correlation function timescales (Appendix \ref{bathcorrelation}). If these conditions are not fulfilled, then the above result are not valid anymore. Instead, when $\tau$ is bigger than the bath correlation timescale, the short time expansion of Eq. \eqref{fseq} changes (Appendix \ref{nmvsm}) and the total QFI becomes bounded by a constant (Appendix \ref{seqmeasurement}) as for Markovian dynamics \cite{pra}.   

 Only one previous work exhibits some similar tendencies for non-Markovian noise effects in harmonic oscillator probe \cite{euroj}.  Here, we show that the results are general properties which do not dependent on the force time modulation neither on the bath spectrum, coupling coefficients nor coupling strength. In \cite{dd} the authors reach similar conclusions for frequency estimation with qubits but treat a special kind of noise which commutes with the parameter encoding transformation.  \\

We show in Appendix \ref{seqmeasurement} that contrary to the single measurement procedure the sequential preparation-and-measurement scenario does not generate loss of information due to an experimental time window $[t_0,t_0+T]$ potentially larger than the one during which the force is actually non-null.

\section{Conclusion}
We have shown the striking difference between Markovian and non-Markovian noise for quantum metrology perspectives, stemming from the ability to perform measurement within the correlation time of the bath, activating non-Markovian effects. Our results finally suggest that we can consider at least three situations. Firstly when measurements are performed within the correlation time of the bath, leading to the super-classical scaling of QFI. Then when measurements are performed on a time scale between the bath correlation time and the time scale $\gamma^{-1}$ emerging from Markovian approximation, leading to the bounded increase of QFI discussed in \cite{pra}. Finally, when measurements are performed on a time scale bigger than $\gamma^{-1}$, leading to no significant increase with respect to the one-measurement strategy, and carrying the whole burden of the noise. The transition between super-classical scaling and bounded QFI should occurs continuously, and should happen when the measurement timescale are comparable with the bath correlation timescale. In  this intermediate situation we cannot conclude about the behaviour of the QFI. \\
One should keep in mind that the ability to perform measurement at time scales shorter than the bath correlation timescale is not enough, it is also necessary that the energy invested in the initial squeezing of the probe state be big enough so that the optimal time interval $\tau_{\mathrm opt}$ goes below the bath correlation timescale. \\
 In \cite{screport,zeno} the authors show an anticorrelation between quantum Zeno effect and Metrological improvement. It is because the estimated parameter depends on the internal dynamics (free evolution of the probe) so that when the evolution is frozen by the quantum Zeno effect, the information about the parameter encoded into the probe is also frozen. In our problem the action of the external force is not affected by the measurements and by the quantum Zeno effect so that the information flow about $F$ from outside into the probe is not frozen.\\
Our method using the global unitary evolution was proved to be efficient to calculate the main dynamical quantities and even the QFI for initial Gaussian states with no other assumptions. The whole difficulty of the exact evolution is concentrated in the bath response function $G(t,t_0)$. It seems that $G(t,t_0)$ captures by itself the nature of the noise, being monotonic for Markovian noise and non-monotonic for resonant narrow band limit. The real behavior of $G(t,t_0)$ is in between. The short time behaviour of $G(t,t_0)$ is also related to quantum Zeno effect. This relationship will be investigated in a future work. \\
We can treat in the same way the probe-bath coupling without rotating wave approximation $(a^{\dag} + a)(B^{\dag}+B)$. The expression of the QFI is also \eqref{firstqfi} but with a more complex function $G(t,u)$. Squeezing effects from the bath appear, turning the identification of the best state and best measurement more difficult. \\
An interesting perspective is to adapt these results to waveform estimation \cite{tsang1,tsang2}.

\begin{acknowledgements}
This  work  is  based  upon  research  supported  by  the
South  African  Research  Chair  Initiative  of  the  Department  of  Science  and  Technology  and  National  Research Foundation. CLL acknowledges and thanks the support of the College of Agriculture, Engineering and Science of the UKZN. 
\end{acknowledgements}

\begin{widetext}
\appendix
\numberwithin{equation}{section}

\section{Operator evolution, first and second moment of probe quadrature operator}\label{evolution}
We first split the operator evolution in a free evolution term and interaction picture operator. Then, we take the probe-bath interaction term apart using the formula 
\bea
&&\exp{\left\{{\cal T}\int_{t_0}^t du [A(u) + B(u)]\right\}} = \exp{\left\{{\cal T}\int_{t_0}^t du \left[e^{{\cal T}\int_{u}^t dsA(s)}B(u)e^{-{\cal A}\int_{u}^t dsA(s)}\right]\right\}}  \exp{\left\{{\cal T}\int_{t_0}^t du A(u)\right\}},
\eea
where ${\cal T}$ is the time ordering operator and ${\cal A}$ is the anti-chronological ordering operator, applying it with $A(u)$ being the probe-bath coupling term and $B(u)$ the force coupling term. It yields
\bea
U(t,t_0,F) &=& e^{-iH_0(t-t_0)/\hbar}\exp{\left\{iF\frac{\omega_0}{\sqrt{2}}{\cal T}\int_{t_0}^t du \zeta(u) \left[ e^{i\omega_0(u-t_0)}{\cal B}(t,u)a^{\dag} {\cal B}^{\dag}(t,u) + c.c.\right]\right\}}{\cal B}(t,t_0), \label{uintermed}
\eea
where $H_0/\hbar = \omega_0 a^{\dag}a + \sum_n \omega_n b_n^{\dag} b_n$,
${\cal B}(t,u) = \exp{{\cal T}\int_u^tds \left[a_0(s)B_0^{\dag}(s) -a_0^{\dag}(s)B_0(s)\right]}$, $a_0(s):=e^{-i\omega_0(s-t_0)}a$ and $B_0(s):=\sum_n K_ne^{-i\omega_n(s-t_0)}b_n$. Then from the Baker-Hausdorff formula one gets
\be\label{btransfo}
{\cal B}(t,u)a^{\dag}{\cal B}^{\dag}(t,u) = a^{\dag}G(t,u) + \int_{u}^tds ~e^{-i\omega_0(s-t_0)}G(s,u)B_0^{\dag}(s),
\ee
with 
\bea\label{gsm}
 &&G(t,u):= 1+ \sum_{k=1}^{\infty}(-1)^k\int_{u}^tds_1\int_{u}^{s_1}ds_2...\int_{u}^{s_{2k-1}}ds_{2k}e^{i\omega_0(s_1-s_2+...+s_{2k-1}-s_{2k})}[B_0(s_1),B_0^{\dag}(s_2)]...[B_0(s_{2k-1}),B_0^{\dag}(s_{2k})]\nn\\
&&= 1+\sum_{k=1}^{\infty}(-1)^k\sum_{n_1,...,n_k}|K_{n_1}|^2...|K_{n_k}|^2\int_{u}^tds_1\int_{u}^{s_1}ds_2...\int_{u}^{s_{2k-1}}ds_{2k} e^{i(\omega_0-\omega_{n_1})(s_1-s_2)}...e^{i(\omega_0-\omega_{n_k})(s_{2k-1}-s_{2k})}.
\eea

One can show that in fact $G(t,u)$ is just a function of $t-u$, $G(t,u)=G(t-u)$. In
the beginning of Appendix \ref{shorttime} we show a integro-differential equation satisfied by $G$ which can be solved by Laplace transform. But in the present problem we only need short time expansion of $G$.\\

Inserting \eqref{btransfo} in \eqref{uintermed}, taking apart the probe terms from the bath terms and calculating the time ordered integral, one obtains, up to an irrelevant phase factor,
\bea\label{evolopapp}
U(t,t_0,F) = &e^{-i\frac{t-t_0}{\hbar}H_0}&e^{iF|D(t,t_0)|X(\phi^D_{t,t_0})} R(t,t_0){\cal B}(t,t_0),
\eea
where $D(t,t_0) = \omega_0\int_{t_0}^tdu \zeta(u) e^{i\omega_0(u-t_0)}G(t,u) $, $\phi^D_{t,t_0}=\arg{D(t,t_0)}$, $X(\phi^D_{t,t_0})=\frac{1}{\sqrt{2}}(a^{\dag}e^{i\phi^D_{t,t_0}} + a e^{-i\phi^D_{t,t_0}})$, and 
\be
R(t,t_0)=e^{iF\frac{\omega_0}{\sqrt{2}}{\cal T}\int_{t_0}^t du \int_u^t ds\zeta(u)
\left[e^{i\omega_0(u-s)}G(s,u)B_0^{\dag}(s) + c.c.\right]}.
\ee

 The time ordered integral can be also calculated in $R(t,t_0)$ and leads to the following expression:
\be
R(t,t_0) = \Pi_n e^{i F |D_n(t,t_0)| X_n(\phi^{D_n}_{t,t_0})}
\ee
with

\be
D_n(t,t_0) = \omega_0 K_n^{*}\int_{t_0}^t du~\zeta(u)e^{i\omega_0(u-t_0)}\int_u^t ds~e^{i(\omega_n-\omega_0)(s-t_0)}G(s,u),
\ee
 $\phi^{D_n}_{t,t_0} = \arg{D_n(t,t_0)}$, and $X_n(\phi^{D_n}_{t,t_0}) = \left(b_n^{\dag}e^{i\phi^{D_n}_{t,t_0}} +b_n e^{-i\phi^{D_n}_{t,t_0}}\right)/\sqrt{2}$.\\

From \eqref{evolopapp} the exact expression in the Heisenberg picture of the probe quadrature $X(\theta)=\frac{1}{\sqrt{2}}(a^{\dag}e^{i\theta}+ae^{-i\theta})$ can be easily derived:
\bea
X(\theta)[t,t_0,F] &=& U_{SB}^{\dag}(t,t_0,F)X(\theta)U_{SB}(t,t_0,F)\nn\\
&=& \frac{1}{\sqrt{2}} {\cal B}^{\dag}(t,t_0)\Bigg\{e^{i\theta}e^{i\omega_0(t-t_0)}\Big[a^{\dag}-i\frac{F}{\sqrt{2}}D^{*}(t,t_0)\Big] + e^{-i\theta}e^{-i\omega_0 (t-t_0)}\Big[a+i\frac{F}{\sqrt{2}}D(t,t_0)\Big]\Bigg\}{\cal B}(t,t_0)\nn\\
&=&  \frac{1}{\sqrt{2}} \Bigg\{e^{i\theta}e^{i\omega_0(t-t_0)}\Big[a^{\dag} G^{*}(t,t_0) -\int_{t_0}^tds e^{-i\omega_0(s-t_0)}G^{*}(t,s)B_0^{\dag}(s)-i\frac{F}{\sqrt{2}}D^{*}(t,t_0)\Big] +c.c.\Bigg\},\label{xquad}
\eea

For the mean value of $X(\theta)[t,t_0,F]$, defined as $\bra X(\theta)[t,t_0,F]\ket = {\rm Tr}_{SB}\{\rho_{SB}^0X(\theta)[t,t_0,F]\}$, we find
\bea
\bra X(\theta)[t,t_0,F]\ket &=& \frac{e^{i\theta}e^{i\omega_0(t-t_0)}}{\sqrt{2}}G^{*}(t,t_0) \bra a^{\dag}\ket_0 +\frac{e^{-i\theta}e^{-i\omega_0(t-t_0)}}{\sqrt{2}} G(t,t_0) \bra a\ket_0  +F|D(t,t_0)|\sin[\theta +\omega_0(t-t_0)-\phi^D_{t,t_0}] \nn\\
 &=& |G(t,t_0)| \big\bra X[\theta+\omega_0(t-t_0) -\phi_{t,t_0}^G]\big\ket_0 +F|D(t,t_0)|\sin[\theta +\omega_0(t-t_0)-\phi^D_{t,t_0}] ,
\eea
where the subscript 0 means that the expectation value is taken for the state $\rho_{SB}^0$. 
We assume also that ${\rm Tr}_{SB}\{\rho_B^0 B_0(s)\}  = {\rm Tr}_{SB}\{\rho_B^0 B_0^{\dag}(s)\} =0$.\\

In order to obtain a simple expression for the variance we make the assumption that the probe and the bath are initially uncorrelated so that $\rho_{SB}^0 = \rho_S^0\rho_B^0$. The variance is 
\bea
\bra \Delta^2 X(\theta)[t,t_0]\ket &=& \Bigg\bra \left[\Delta\left(\frac{1}{\sqrt{2}}e^{i\theta + i\omega_0(t-t_0)} G^{*}(t,t_0)a^{\dag} + c.c.\right)\right]^2\Bigg\ket_0 \nn\\
&+& \int_{t_0}^tds\int^t_{t_0}ds' G^{*}(t,s') G(t,s)\sum_n |K_n|^2 e^{i(\omega_0-\omega_n)(s-s')}\left(N_n + \frac{1}{2}\right)\nn\\
 &=& |G(t,t_0)|^2\Big\bra \Delta^2 X\big[\theta +\omega_0(t-t_0)-\phi^G_{t,t_0}\big]\Big\ket_0 +\sum_n |K_n|^2\left(N_n + \frac{1}{2}\right) \Bigg|\int_{t_0}^tds G(t,s) e^{i(\omega_0-\omega_n)s}\Bigg|^2\nn\\\label{genvariance}
\eea
with $N_n := \bra b_n^{\dag}b_n\ket$ and $\phi^G_{t,t_0}:=\arg G(t,t_0)$. Since the second term in \eqref{genvariance} does not depend on initial conditions, the variance $\bra \Delta^2 X(\theta)[t,t_0]\ket $ is maximal whenever $\Big\bra \Delta^2 X\big[\theta +\omega_0(t-t_0)-\phi^G_{t,t_0}\big]\Big\ket_0 $  is maximal. This gives the relation between $\theta_m^t$ the angle of the maximal quadrature variance at time $t$, and $\theta_m^0$ the angle of the maximal quadrature variance at time $t_0$: 
\be
\theta_m^t = \theta_m^0 + \phi_{t,t_0}^G -\omega_0(t-t_0).
\ee

The variance of $P(\phi^D_{t,t_0}-\omega_0(t-t_0))[t,t_0]$ is obtained doing $\theta=\phi^D_{t,t_0}-\omega_0(t-t_0) + \pi/2$ in \eqref{genvariance}:
\be\label{variancepbest}
\big\bra \Delta^2 P(\phi^D_{t,t_0}-\omega_0(t-t_0))[t,t_0]\big\ket = |G(t,t_0)|^2\big\bra \Delta^2 P\big[\phi^D_{t,t_0}-\phi^G_{t,t_0}\big]\big\ket_0 +\sum_n |K_n|^2\left(N_n + \frac{1}{2}\right) \Bigg|\int_{t_0}^tds G(t,s) e^{i(\omega_0-\omega_n)s}\Bigg|^2,\nn\\
\ee
and is minimized when $\big\bra \Delta^2 P\big[\phi^D_{t,t_0}-\phi^G_{t,t_0}\big]\big\ket_0$ is minimized. According to \cite{pra}, the state that realizes this minimization, given an initial mean energy $E$, is a pure squeezed state with the squeezed quadrature being $P\big[\phi^D_{t,t_0}-\phi^G_{t,t_0}\big]$. It can be written as $\hat S[\mu(t,t_0)]|0\rangle$ where $|0\rangle$ is the ground state of the harmonic oscillator, $\hat S[\mu(t,t_0)] = \exp{\left(\frac{\mu(t,t_0)}{2} a^{\dag 2} - \frac{\mu^{*}(t,t_0)}{2} a^{2}\right)}$ with $\mu(t,t_0)=re^{2i(\phi^D_{t,t_0}-\phi^G_{t,t_0})}$ and $r=\frac{1}{2}\ln{[2(E+\sqrt{E^2-1/4})]}$. The corresponding minimal variance is:
\be
\bra 0|\hat S^{\dag}[\mu(t,t_0)] \left[\Delta P(\phi^D_{t,t_0}-\phi^G_{t,t_0})\right]^2\hat S[\mu(t,t_0)]|0\rangle =\frac{1}{4} \left(E+\sqrt{E^2-1/4}\right)^{-1} ,
\ee
and
\be
\bra 0|\hat S^{\dag}[\mu(t,t_0)] \left[\Delta X(\phi^D_{t,t_0}-\phi^G_{t,t_0})\right]^2\hat S[\mu(t,t_0)]|0\rangle = \left(E+\sqrt{E^2-1/4}\right) ,
\ee
is maximal.\\

The determinant of the covariance matrix ${\rm Det}{\bf \Sigma}(t,t_0)$ can be written for any quadrature $X(\theta)[t,t_0]$ and $P(\theta)[t,t_0]$ as:
\be
{\rm Det}{\bf \Sigma}(t,t_0) = \bra \{\Delta X(\theta)[t,t_0]\}^2\ket_0 \bra\{\Delta P(\theta)[t,t_0]\}^2\ket_0 - \frac{1}{4}\Big\bra \Delta X(\theta)[t,t_0]\Delta P(\theta)[t,t_0] +\Delta P(\theta)[t,t_0]\Delta X(\theta)[t,t_0]\Big\ket_0^2.
\ee 

Using \eqref{xquad}, \eqref{genvariance} and conjugated expressions one finds
\be
{\rm Det}{\bf \Sigma}(t,t_0) = |G(t,t_0)|^4{\rm Det}{\bf \Sigma}^0 + |G(t,t_0)|^2\langle \Delta a\Delta a^{\dag} +\Delta a^{\dag} \Delta a\rangle_0 n_B(t,t_0) + n_B^2(t,t_0),
\ee
where ${\bf \Sigma}^0$ is the initial covariance matrix of $S$, $\Delta a := a- \langle a\rangle_0$, and the noise term from the bath is noted $n_B(t,t_0) : = \sum_n |K_n|^2\left(N_n + \frac{1}{2}\right) \Big|\int_{t_0}^tds G(t,s) e^{i(\omega_0-\omega_n)s}\Big|^2$. As expected ${\rm Det}{\bf \Sigma}(t,t_0)$ does not depend on $\theta$ and is minimal when $S$ is initialized in a pure state. \\

\section{Narrow band limit, broad band limit and Markovian dynamics}\label{bnband}
The narrow band limit is obtained for instance retaining only one mode of the bath and taking to zero the coupling coefficient of the other modes. One can also convert the discrete bath to a continuum of mode and take the mode distribution to a delta function. In any case we are left with one mode of frequency $\omega$ and coupling coefficient $K_0$. The function $G(t,t_0)$ simplifies to:
\bea
 &&G(t,t_0)= 1+\sum_{k=1}^{\infty}(-1)^k|K_0|^{2k}\int_{u}^tds_1\int_{t_0}^{s_1}ds_2...\int_{t_0}^{s_{2k-1}}ds_{2k} e^{i(\omega_0-\omega_{n_1})(s_1-s_2...+s_{2k-1}-s_{2k})},
\eea
and if we assume also that the bath mode is resonant with the probe,
\bea
 G(t,t_0)&=& 1+\sum_{k=1}^{\infty}(-1)^k|K_0|^{2k}\int_{u}^tds_1\int_{t_0}^{s_1}ds_2...\int_{t_0}^{s_{2k-1}}ds_{2k} 1\nn\\
&=& 1+\sum_{k=1}^{\infty}(-1)^k|K_0|^{2k}\frac{1}{2k!}(t-t_0)^{2k}\nn\\
&=&\cos{|K_0|(t-t_0)}.
\eea

We recover a sinusoidal behavior, meaning that the information is just going to the bath and coming back entirely to the probe periodically. \\

We take now the broad band limit. As we will see one needs more assumptions to recover the Markovian dynamics described in \cite{pra}. \\

Firstly the bath modes are taken to form a continuum,
\be 
\sum_n |K_n|^2 \rightarrow \int_0^{\omega_c} d\omega g(\omega)|K(\omega)|^2 ,
\ee
where $g(\omega)$ is the bath mode distribution and $\omega_c$ is a cut-off (for instance one can consider that the experiment takes place in a limited volume). We now assume that $g(\omega)$ and $K(\omega)$ are mode independent. We have then
\bea
[B_0(s_{2k-1}),B_0^{\dag}(s_{2k})] &=& \sum_n |K_n|^2 e^{-i\omega_n(s_{2k-1}-s_{2k})}\nn\\
&=& \int_0^{\omega_c}d\omega g(\omega)|K(\omega)|^2e^{-i\omega_n(s_{2k-1}-s_{2k})}\nn\\
&=& g|K|^2 \int_0^{\omega_c}d\omega e^{-i\omega_n(s_{2k-1}-s_{2k})}\nn\\
&=&g|K|^2 \left[\pi \delta(s_{2k-1}-s_{2k}) -i {\cal P}\frac{1}{s_{2k-1}-s_{2k}}\right],\label{pb}
\eea
with the third line corresponding to mode-independent coupling coefficients and spectrum density, and the fourth line is the limit of $\omega_c$ going to infinity. The Markovian dynamics corresponds to the real part of \eqref{pb}. The imaginary part will generate non-convergent terms when integrating $s_{2k}$ from $u$ to $s_{2k-1}$. So one should maintain the cut-off $\omega_c$ finite, and face the subsequent integrations. To recover the Markovian dynamics one need one more assumption in order to be able to discard the imaginary part. This assumption is equivalent to the rotating wave approximation. Instead of integrating the mode form 0 to $\omega_c$ we integrate from $-\omega_c$ to $\omega_c$ with $g(-\omega)=g(\omega)$ and $|K(-\omega)|=|K(\omega)|$. Then the imaginary part of \eqref{pb} cancels out and we end up with $[B_0(s_{2k-1}),B_0^{\dag}(s_{2k})] =g|K|^2 \pi \delta(s_{2k-1}-s_{2k})$. The integration from $-\omega_c$ to $\omega_c$ can be justified through the rotating wave approximation: the negative frequencies are far from resonance and thus contribute very little as soon as $t-t_0 \gg (\omega_0 -\omega)^{-1}$ (with $\omega \in [-\omega_c;0]$) \cite{gardiner}. We get from this that we can legitimately discard the imaginary part of \eqref{pb} as soon as we are interested in times bigger than $\omega_0^{-1}$.  \\

We show now that substituting only the real part of \eqref{pb} in the expression of $G(t,t_0)$ we recover the Markovian behavior \cite{pra}:
\bea
G(t,t_0) &=& 1 +\sum_{k=1}^{\infty}(-1)^k \frac{\pi^k}{2^k}g^k|K|^{2k}\int_{t_0}^t ds_1\int_{t_0}^{s_1}ds_3...\int_{t_0}^{s_{2k-3}}ds_{2k-1}\nn\\
&=& 1+\sum_{k=1}^{\infty}\frac{(-\pi g|K|^2/2)^k}{k!}(t-t_0)^k\nn\\
&=& e^{-\gamma(t-t_0)/2},
\eea
where $\gamma = \pi g |K|^2$. Applying this result to the variance of $P(\theta)$ one find
\bea
\bra \Delta^2 P(\theta)[t,t_0]\ket &=& e^{-\gamma(t-t_0)}\Bigg\bra \left[\Delta\left(\frac{i}{\sqrt{2}}e^{i\theta + i\omega_0(t-t_0)} a^{\dag} - c.c.\right)\right]^2\Bigg\ket_0 +\sum_n |K_n|^2\left(N_n + \frac{1}{2}\right) \Bigg|\int_{t_0}^tds e^{-\gamma(t-s)/2} e^{i(\omega_0-\omega_n)s}\Bigg|^2\nn\\
&=& e^{-\gamma(t-t_0)}\Bigg\bra \left[\Delta\left(\frac{i}{\sqrt{2}}e^{i\theta + i\omega_0(t-t_0)} a^{\dag} - c.c.\right)\right]^2\Bigg\ket_0 \nn\\
&&+\int_0^{\infty}d\omega g(\omega)|K(\omega)|^2\left(N(\omega) + \frac{1}{2}\right) \int_{t_0}^tds \int_{t_0}^t ds'e^{-\gamma(2t-s-s')/2} e^{i(\omega_0-\omega_n)(s-s')}\nn\\
&=& e^{-\gamma(t-t_0)}\big\bra \left\{\Delta P[\theta +\omega(t-t_0)]\right\}^2\big\ket_0 +g|K|^2\left(N + \frac{1}{2}\right) \int_{t_0}^tds \int_{t_0}^t ds'e^{-\gamma(2t-s-s')} \pi\delta(s-s')\nn\\
&=& e^{-\gamma(t-t_0)}\big\bra \left\{\Delta P[\theta +\omega(t-t_0)]\right\}^2\big\ket_0 +\left(N + \frac{1}{2}\right) \left(1-e^{-\gamma(t-t_0)}\right),
\eea
where we also assume that the mean excitation number $N(\omega)$ is mode-independent $N(\omega)=N$.  \\

The Fisher information from the best quadrature measurement becomes 
\be
{\cal F}_{P(\phi^D_{t,t_0} -\omega_0(t-t_0))}(t,t_0) = \frac{\omega_0^2\big|\int_{t_0}^tdu \zeta(u) e^{i\omega_0(u-t_0)}e^{-\gamma(t-u)/2}\big|^2}{\bra \Delta^2 P(\phi^D_{t,t_0} -\omega_0(t-t_0))[t,t_0]\ket}=\frac{\omega_0^2\big|\int_{t_0}^tdu \zeta(u) e^{i\omega_0(u-t_0)}e^{-\gamma(t-u)/2}\big|^2}{e^{-\gamma(t-t_0)}\big\bra \left\{\Delta P(\phi^D_{t,t_0})\right\}^2\big\ket_0 +\left(N + \frac{1}{2}\right) \left(1-e^{-\gamma(t-t_0)}\right)},\label{markovqfi}
\ee
which is precisely the expression found in \cite{pra} for a Markovian dynamics.\\

In conclusion, the Markovian approximation is more than just the broad band limit, it is also the mode-independence of the coupling coefficients and spectrum but also the rotating wave approximation. This approximation is valid as soon as we are interested in times $t-t_0$ much bigger than $\omega_0^{-1}$. This contributes to the fact that in general Markovian approximation fails to describe the real dynamics at short times.

\section{Quadrature measurement}\label{quadmeas}
We consider the measurement of the quadrature $P(\theta) = \frac{i}{\sqrt{2}}(e^{i\theta}a^{\dag} - e^{-i\theta}a)$. The output is $p$ and the probe is projected on the eigenstate $|p,\theta\ket$ such that $P(\theta) |p,\theta\ket =p |p,\theta\ket$. One POVM corresponding to such an ideal projective measurement is $\{|p,\theta\ket\bra p,\theta|\}_{p \in {\mathcal R}}$. The output conditional distribution is ${\cal P}(p,\theta|F) = {\rm Tr}_S\{\rho_s(t,t_0,F)|p,\theta\ket \bra p,\theta|\}$, where $\rho_S(t,t_0,F)={\rm Tr}_B\{U(t,t_0,F)\rho_{SB}^0U^{\dag}(t,t_0,F)\}$ is the state of the probe at the instant t after interacting with the force and the bath since the instant $t_0$ and when the value of the force amplitude is $F$. The eigenstates of $P(\theta)$ cannot be normalized so that the direct calculation of ${\cal P}(p,\theta|F)$ is not easy. But the global Hamiltonian is quadratic in the probe and bath operators so that the global evolution is a Gaussian operation. So if the initial state $\rho_{SB}^0$ is a Gaussian state the final state is Gaussian too. The reduced state of $S$ is also Gaussian (partial tracing is a Gaussian operation), and it is characterized only by the average $\bra P(\theta)[t,t_0,F]\ket := {\rm Tr}_S\{\rho_S(t,t_0,F)P(\theta)\}$ and the variance $\bra \Delta^2 P(\theta)[t,t_0]\ket := {\rm Tr}_S\{\rho_S(t,t_0,F)[P(\theta)-\bra P(\theta)[t]\ket]^2\}$:

\bea
{\cal P}(p,\theta|F) &=& \frac{1}{2\pi \big\bra \Delta^2 P(\theta)[t,t_0]\big\ket}\exp{\Big\{-\frac{1}{2\big\bra \Delta^2 P(\theta)[t,t_0]\big\ket}\big\{p - \big\bra  P(\theta)[t,t_0,F]\big\ket\big\}^2\Big\}}\nn\\
&=& \frac{1}{2\pi \big\bra \Delta^2 P(\theta)[t,t_0]\big\ket}\nn\\
&&\!\!\!\!\!\!\!\!\!\!  \!\!\!\!\!\!\!\!\!\!\!\!\!\!\!\!\!\!\!\! \!\!\! \times\exp{\Bigg\{-\frac{\Big\{p -\frac{ie^{i\theta}}{\sqrt{2}} e^{i\omega_0(t-t_0)}G^{*}(t,t_0) \bra a^{\dag}\ket_0 +\frac{ie^{-i\theta}}{\sqrt{2}}e^{-i\omega_0(t-t_0)}G(t,t_0) \bra a\ket_0  -F|D(t,t_0)|\cos[\theta +\omega_0(t-t_0)-\phi^D_{t,t_0}] \Big\}^2}{2\big\bra \Delta^2 P(\theta)[t,t_0]\big\ket}\Bigg\}}.\nn\\
\eea
The derivation of the expression of $\bra P(\theta)[t,t_0,F]\ket$ is shown in Appendix \ref{evolution}.\\

When the parameter to estimate is described by a Gaussian distribution the Fisher information is equal to the inverse of its variance. We are presently in such a situation and it is clear that the variance of the distribution ${\cal P}(p,\theta|F) $ seen as a distribution of the parameter $F$ is $\bra \Delta^2 P(\theta)[t,t_0]\ket\left\{|D(t,t_0)|^2\cos^2\left[\theta +\omega_0(t-t_0)-\phi^D_{t,t_0}\right]\right\}^{-1}$.\\

So the Fisher information corresponding to the measurement of $P(\theta)$ is 
\be
{\cal F}_{P(\theta)}(t,t_0) = \frac{|D(t,t_0)|^2\cos^2[\theta +\omega_0(t-t_0)-\phi^D_{t,t_0}]}{\bra \Delta^2 P(\theta)[t,t_0]\ket}.
\ee

One can easily see that the best quadrature measurement  is for the angle $\theta = \phi^D_{t,t_0} -\omega_0(t-t_0)$ such that $\cos^2[\theta +\omega_0(t-t_0)-\phi^D_{t,t_0}] =1$ so that the Fisher information from the best quadrature measurement is 
\bea\label{fibestmeasurement}
{\cal F}_{P(\phi^D_{t,t_0} -\omega_0(t-t_0))}(t,t_0) =\frac{|D(t,t_0)|^2}{\big\bra \Delta^2 P(\phi^D_{t,t_0} -\omega_0(t-t_0))[t,t_0]\big\ket}
\eea
and it is exactly the expression of the quantum Fisher information in Eq. (6) of the main text.

\section{Short time behavior}\label{shorttime}
\subsection{Defining the short time regime}\label{considtimescale}
One can show that the function $G(t,t_0)$ satisfies the following integro-differential equation:
\be\label{eqdiff}
\dot{G}(t,t_0):=\frac{d}{dt}G(t,t_0) = -\int_{t_0}^tds\sum_n|K_n|^2e^{i(\omega_0-\omega_n)(t-s)}G(s,t_0).
\ee

One could solve this equation by Laplace transform but, here, we only need the short time behaviour.\\

From this relation one can derive the successive derivatives evaluated in $t=t_0$ for any integer $p\geq2$,
\be
\frac{d^p}{dt^p}G(t,t_0)_{|_{t=t_0}} =- \sum_n|K_n|^2\sum_{l=0}^{p-2}i^{p-2-l}(\omega_0-\omega_n)^{p-2-l}\frac{d^{l}}{dt^{l}}G(0),
\ee 
with the notation $\frac{d^{l}}{dt^{l}}G(0):=\frac{d^l}{dt^l}G(t,t_0)_{|_{t=t_0}}$, making explicit the fact that $G(t,t_0)$ is a simple function of $t-t_0$ as mentioned in Appendix \ref{evolution}. For $p=0$ and $p=1$ we have $G(0)=1$ and $\frac{d}{dt}G(0)=0$. Consequently, the successive derivatives of $G(t,t_0)$ are sums and products of terms $\sum_n|K_n|^2i^l(\omega_0-\omega_n)^l$, with the powers of the $|K_n|$-factors and $(\omega_0-\omega_n)$-factors summing up to the order of the derivative. One can conclude that the evolution time scales of $G(t,t_0)$ are of the order $\Omega_p^{-1}$, with $\Omega_p$ defined for all $p\geq 2$ as
\be
\Omega_p:=\left|\sum_n|K_n|^2(\omega_0-\omega_n)^{p-2}\right|^{1/p}.
\ee

 This gives us a condition for the validity of the expansion of $G(t,t_0)$ in $t=t_0$, so that when $t-t_0$ is much smaller than $\Omega_p^{-1}$, $\forall p\geq 2$, one can retain the first terms of this expansion:
\be\label{gexp}
G(t,t_0)= 1 -\frac{{\cal K}^2}{2}(t-t_0)^2+i\frac{(t-t_0)^3}{6}\sum_n|K_n|^2(\omega_n-\omega_0) + {\cal O}[\Omega_4(t-t_0)^4],
\ee
where ${\cal K}^2:=\sum_n|K_n|^2$.\\

Note that a slow time scale $\gamma^{-1}$ can also emerge for long times (Appendix \ref{longtime}) or Markovian approximation (Appendix \ref{bnband}).\\

One can expand $D(t,t_0)$ under the same conditions, but adding also the conditions $t-t_0$ much smaller than $\omega_0^{-1}$ and the evolution time scale of $\zeta(t)$:
\be
D(t,t_0)=\omega_0 (t-t_0)\zeta(t_0) + \omega_0\frac{(t-t_0)^2}{2}[i\omega_0\zeta(t_0)+\dot{\zeta}(t_0)] + {\cal O}[\omega_0{\cal K}^2(t-t_0)^3].
\ee

Finally, in order to expand \eqref{qfipvar}, we have to consider also the expansion of $\big\bra \Delta^2 P(\phi^D_{t,t_0} -\omega_0(t-t_0))[t,t_0]\big\ket$. From Eq. \eqref{variancepbest} we already have an expression available, but we will use the following form in order to simplify the considerations on time scales: 
\be
\big\bra \Delta^2 P(\phi^D_{t,t_0} -\omega_0(t-t_0))[t,t_0]\big\ket = |G(t,t_0)|^2\Big\bra \Delta^2 P\big[\phi^D_{t,t_0}-\phi^G_{t,t_0}\big]\Big\ket_0  + \int_{t_0}^tds\int_{t_0}^tds' G(t,s)G^{*}(t,s')C^0(s-s'),\label{newvar}
\ee
where 
\be
C^0(s-s'):=e^{i\omega_0(s-s')}\frac{1}{2}{\rm Tr}_B\{\rho_B^0[B_0(s)B_0^{\dag}(s') + B_0^{\dag}(s')B_0(s)]\}=\sum_n|K_n|^2(N_n+1/2)e^{i(\omega_0-\omega_n)(s-s')}.
\ee

To be allowed to take into account only the first terms of the short time expansion of  $\big\bra \Delta^2 P(\phi^D_{t,t_0} -\omega_0(t-t_0))[t,t_0]\big\ket $, one has to consider times $t-t_0$ much smaller than the $\Omega_p^{-1}$ but also much smaller than the evolution time scale of $C^0(s'-s)$. Note that although having explicit time dependence within the phase of the quadrature variance, the actual value of $\Big\bra \Delta^2 P\big[\phi^D_{t,t_0}-\phi^G_{t,t_0}\big]\Big\ket_0$ depends only on the initial probe state. Nevertheless, we do not need this argument. From the short time expansion of $D(t,t_0)$ and $G(t,t_0)$ one can see that the bath contribution appears only at the third order in $(t-t_0)$ for $\phi^G_{t,t_0} = \arg{G(t,t_0)}$ and second order in $(t-t_0)$ for $\phi^D_{t,t_0}=\arg{D(t,t_0)}$. Then the variance can be expanded in the following way:
\be
\Big\bra \Delta^2 P\big[\phi^D_{t,t_0}-\phi^G_{t,t_0}\big]\Big\ket_0 = \Big\bra \Delta^2 P\big[\phi^{D_0}_{ t,t_0}\big]\Big\ket_0 + {\cal O}({\cal K}^2(t-t_0)^2),
\ee
where 
\be
D_0(t,t_0) := \omega_0\int_{t_0}^tdu \zeta(u) e^{i\omega_0(u-t_0)} ,
\ee
is the coefficient $D(t,t_0)$ in the noiseless situation ($K_n \rightarrow 0$ $\forall n$). \\

Regarding the second term in Eq. \eqref{newvar}, one can easily re-write the Taylor expansion of $C^0(s'-s)$ in 0 in the following form
\be
C^0(s-s') = \sum_n |K_n|^2(N_n+1/2)\left[1+ \sum_{p=1}^{\infty} \frac{[i\chi_p(s-s')]^p}{p!}\right],
\ee
where the frequencies defining the evolution time scales $|\chi_p|^{-1}$ are given by $\chi_p := \left[\sum_n\frac{|K_n|^2(N_n+1/2)}{\cal N}(\omega_0-\omega_n)^p\right]^{1/p}$ and ${\cal N}:=\sum|K_n|^2(N_n+1/2)$. Note that if the $\Omega_p$ converge then the $\chi_p$ too, since the prefactor $\frac{|K_n|^2(N_n+1/2)}{\cal N}$ within the sum is smaller than $|K_n|^2$. \\

Recapping the above considerations on time scales, as long as $(t-t_0)$ is smaller than $\omega_0^{-1}$, $|\chi_q|^{-1}$, $q\geq1$, and $\Omega_p^{-1}$, $p\geq2$, we can write
\be
\big\bra \Delta^2 P(\phi^D_{t,t_0} -\omega_0(t-t_0))[t,t_0]\big\ket  = \Big\bra \Delta^2 P\big[\phi^{D_0}_{ t,t_0}\big]\Big\ket_0 + {\cal O}[({\cal K}^2+{\cal N}^2)(t-t_0)^2],
\ee
and one can also expand $G(t,t_0)$, $D(t,t_0)$, $\big\bra \Delta^2 P(\phi^D_{t,t_0} -\omega_0(t-t_0))[t,t_0]\big\ket$ and the QFI (Eq. \eqref{qfipvar}), and finally get
\bea
{\cal F}_{P(\phi^D_{t,t_0} -\omega_0(t-t_0))}(t,t_0) = \frac{\omega_0^2}{\big\bra \Delta^2 P\big[\phi_{t,t_0}^{D_0}\big]\big\ket_0 }\{\zeta^2(t_0)(t-t_0)2 +\zeta(t_0)\dot{\zeta}(t_0)(t-t_0)^3 + {\cal O}[\omega_0^2{\cal K}^2(t-t_0)^4]\}.\label{stexp}
\eea

\subsection{Non-Markovian effects at short times}\label{nmvsm}
In the above subsection we derive conditions of validity of the short time expansion \eqref{stexp} of the QFI. We show in Appendix \ref{bathcorrelation} that these time scales correspond also to the evolution time scales of the bath correlation function. If those conditions on $t-t_0$ cannot be fulfilled, meaning that the bath correlation time is not accessible and no measurement can be performed below the bath correlation time, the higher orders of the expansion \eqref{stexp} cannot be neglected and the first terms are not significant any more. Then the correct expansion is not anymore centered at $t_0$ but at $t_0+t_c$, where $t_c$ represents the bath correlation time. In such an expansion, the first derivative of $G$ is taken at $t_0+t_c$ and does not cancel anymore, yielding a first order bath dependent term for the short time expansion of $G(t,t_0)$ and of the denominator of \eqref{qfipvar}, as for Markovian dynamics (Appendix \ref{bnband}). As a comparison, Markovian dynamics can be sketched as the impossibility of accessing the correlation time of the bath and then considering that $t_c$ goes to zero. We show at the end of Appendix \ref{seqmeasurement} that the presence of a first order term in the denominator of \eqref{qfipvar} changes dramatically the QFI behaviour for sequential preparation-and-measurement scenario. The QFI becomes bounded by a constant, irrespectively of the energy invested in the squeezing of the probe initial state.\\

As a matter of comparison, we give here the short time expansion of \eqref{qfipvar} for Markovian dynamics (obtained from Eq. \eqref{markovqfi}) valid for $t-t_0$ smaller than $\omega_0^{-1}$, $\gamma^{-1}$ (defined in \ref{bnband}) and the evolution time scale of $\zeta(t)$:
\bea
{\cal F}^M_Q(t,t_0) = &\frac{\omega_0^2(t-t_0)^2}{\big\bra \Delta^2 P\big[\phi_{t,t_0}^D\big]\big\ket_0 }&\left\{\zeta^2(t_0)+\left[\zeta(t_0)\dot{\zeta}(t_0)+\frac{\gamma}{2}\zeta^2(t_0)-\frac{\gamma\left(n_T+\frac{1}{2}\right)}{\big\bra \Delta^2 P\big[\phi_{t,t_0}^D\big]\big\ket_0} \zeta^2(t_0)\right](t-t_0)\right\}\nn\\
&+& {\cal O}[\omega_0^2(t-t_0)^2].
\eea

The bath contribution comes also at the 3rd order and always reduces the amount of information since $\gamma/2-\gamma\left(n_T+\frac{1}{2}\right) \Big\bra \Big\{\Delta P\big[\phi_{t,t_0}^D\big]\Big\}^2\Big\ket_0^{-1}$ is always strictly negative (even smaller than $-\gamma/2$). This happens because the only contribution in the first derivative of $G(t,t_0)$ at $t=t_0$ comes from $-\sum_n|K_n|^2\int_{t_0}^t ds e^{i(\omega_0-\omega_n)(t-s)}$ (see Eq. \eqref{eqdiff}). If one takes the broad band limit together with the rotating wave approximation one ends up with $-\gamma/2$  (see Appendix \ref{bnband}). This implies that the short time behavior of the QFI for the Markovian dynamics is qualitatively different as we saw above.\\

 Hence the apparition of the bath contribution only at the 4th order is a particularity of the expansion \eqref{stexp} and comes from measurements at time scales below the correlation time of the bath, justifying its classification as non-Markovian effects.

\section{Long time behavior}\label{longtime}
We are interested in the long time behavior of ${\cal G}_1(t,t_0)$ the first term of the sum in the expression of $G(t,t_0)$ \eqref{gsm}.
\bea
{\cal G}_1(t,t_0) &:=& - \int_{t_0}^tds_1\int_{t_0}^{s_1}ds_2e^{i\omega_0(s_1-s_2)}\left[B_0(s_1),B_0^{\dag}(s_2)\right]\label{g1}\\
&=& -\sum_n |K_n|^2 \left[-\frac{i(t-t_0)}{\omega_n-\omega_0} +\frac{1-e^{-i(\omega_n -\omega_0)(t-t_0)}}{(\omega_n-\omega_0)^2}\right]\nn\\
&=& - \sum_n |K_n|^2 \left[\frac{1-\cos{\{(\omega_n -\omega_0)(t-t_0)\}}}{(\omega_n-\omega_0)^2} -i\frac{t-t_0}{\omega_n-\omega_0}+i\frac{\sin{\{(\omega_n -\omega_0)(t-t_0)\}}}{(\omega_n-\omega_0)^2}\right].
\eea

One can show that when $t-t_0$ goes to infinity the real part of the integrand tends to 
\be
\frac{1-\cos{\{(\omega_n -\omega_0)(t-t_0)\}}}{(\omega_n-\omega_0)^2} \rightarrow \frac{\pi}{2} (t-t_0)\delta(\omega-\omega_0).
\ee

The long time behavior of the real part of ${\cal G}_1(t,t_0)$ reproduces the Markovian behavior since we recover a $\delta-function$. Substituting in the expression of ${\cal G}_1(t,t_0)$ we find
\bea
\Re{{\cal G}_1(t,t_0)} &=& -\sum_n |K_n|^2 \frac{1-\cos{\{(\omega_n -\omega_0)(t-t_0)\}}}{(\omega_n-\omega_0)^2}\nn\\
&=& -\int_0^{\infty}d\omega g(\omega)|K(\omega)|^2 \frac{1-\cos{\{(\omega_n -\omega_0)(t-t_0)\}}}{(\omega_n-\omega_0)^2}\nn\\
&&\rightarrow -\frac{\pi}{2}(t-t_0)g(\omega_0)|K(\omega_0)|^2.
\eea

In the second line we substitute the discrete bath mode distribution by a continuous one in order to realize the integration. 
Note that the Markovian limit gives a similar result ${\cal G}_1(t,t_0)\rightarrow -\gamma(t-t_0)/2=-\pi(t-t_0)g|K|^2/2$. So for the real part of ${\cal G}_1(t,t_0)$ the long time limit is similar to the Markov approximation. However, it is not so simple for the imaginary part, the same treatment as for the real part leads to an undetermined form. Writing the sine of the imaginary part in a series expansion one obtains the following expression:
\bea
\Im{{\cal G}_1(t,t_0)} =\sum_{p=0}^{\infty}(-1)^{p+1}\frac{(t-t_0)^{2p+3}}{(2p+3)!}\big\langle(\omega_0-\omega)^{2p+1}\big\rangle,
\eea
where $\big\langle(\omega_0-\omega)^{2p+1}\big\rangle = \int_0^{+\infty} d\omega g(\omega)|K(\omega)|^2(\omega_0-\omega)^{2p+1}$. The sum is expected to converge since the imaginary part $\Im{{\cal G}_1(t,t_0)}$ is finite (can be seen form expression \eqref{g1}). Note that if $g(\omega)|K(\omega)|^2$ is symmetrical with respect to $\omega_0$ the imaginary part $\Im{{\cal G}_1(t,t_0)}$ cancels out.

\section{Sequential preparation-and-measurement scenario}\label{seqmeasurement}
We analyze the variance of the quadrature $P(\phi^D_{t_{k+1},t_k} -\omega_0\tau)$ after the probe interacting with the force and the bath from $t_k:=t_0+k\tau$ to $t_{k+1}:=t_0+(k+1)\tau$. We use the expression of the variance \eqref{newvar} used in Appendix \ref{considtimescale}
\bea
\bra \Delta^2 P(\phi^D_{t_{k+1},t_k} -\omega_0\tau)[t_{k+1},t_k]\ket &=&|G(t_{k+1},t_k)|^2\Big\bra \Delta^2 X\big[\phi^D_{t_{k+1},t_k}-\phi^G_{t_{k+1},t_k}\big]\Big\ket_0 \nn\\
& +& \int_{t_k}^{t_{k+1}}ds\int_{t_k}^{t_{k+1}}ds' G(t_{k+1},s)G^{*}(t_{k+1},s')C^0(s-s').\label{seqvar}
\eea

One can show that $G(t,u) =G(t-u)$, yielding $G(t_{k+1},t_k) = G(\tau)$ and that the double integral depends only on $t_{k+1}-t_k$, allowing to re-write \eqref{seqvar} as
\bea
\bra \Delta^2 P(\phi^D_{t_{k+1},t_k} -\omega_0\tau)[t_{k+1},t_k]\ket &=&|G(\tau)|^2\Big\bra \Delta^2 X\big[\phi^D_{t_{k+1},t_k}-\phi^G_{\tau}\big]\Big\ket_0 +\int_0^{\tau}ds\int_0^{\tau}ds' G(\tau-s)G^{*}(\tau-s')C^0(s-s'),\nn\\
\eea
where $\phi^G_\tau :=\arg{G(\tau)}$. As discussed in Section V of the main text we make the assumption that the probe is prepared in the best state $\hat S[\mu(t_{k+1},t_k)]|0\rangle$ after each measurement. Thanks to this assumption, which corresponds to the best strategy, the expression of the variance simplifies to
\bea\label{varpappendix}
\bra \Delta^2 P(\phi_{t_{k+1},t_k} -\omega_0\tau)[t_{k+1},t_k]\ket &=&\frac{1}{4} \big|G(\tau)\Big|^2\left(E+\sqrt{E^2-1/4}\right)^{-1} +\int_0^{\tau}ds\int_0^{\tau}ds' G(\tau-s)G^{*}(\tau-s')C^0(s-s'),\nn\\
\eea
since $\bra 0|\hat S^{\dag}[\mu(t_{k+1},t_k)] \left[\Delta P(\phi^D_{t_{k+1},t_k}-\phi^G_\tau)\right]^2\hat S[\mu(t_{k+1},t_k)]|0\rangle =\frac{1}{4} \left(E+\sqrt{E^2-1/4}\right)^{-1} $.
The expression \eqref{varpappendix} depends only on $\tau$ and not anymore on k so that the denominator in Eq. \eqref{fseq}  can be taken out of the sum.\\

Assuming now that $\tau$ is much smaller than all time scales involved $D(t_k+\tau,t_k)$, that is much smaller than $\omega_0^{-1}$, $\Omega_p^{-1}$, $p\geq2$ (see Appendix \ref{considtimescale}), and the evolution time scale of $\zeta(t)$, we can expand $D(t_k+\tau,t_k)$ to order 2:
\be
D(t_k+\tau, t_k) = \omega_0\tau\zeta(t_k)+ \omega_0\frac{\tau^2}{2}\left[\dot{\zeta}(t_k) + i\omega_0\zeta(t_k) \right]+ {\cal O}(\tau^3),
\ee 
where the dot means the time derivative. For $|D(t_k,t_k+\tau)|^2$ we have:
\be\label{exp}
|D(t_k,t_k+\tau)|^2 = \omega_0^2 \tau^2\left[\zeta^2(t_k)\left(1+\omega_0^2\frac{\tau^2}{4}\right) +\tau\zeta(t_k)\dot{\zeta}(t_k)+ \frac{\tau^2}{4}\dot{\zeta}^2(t_k) \right]+ {\cal O}(\tau^5).
\ee
 
The Euler-Maclaurin formula relates the sum $\sum_{k=0}^{k=\nu-1}|D(t_k,t_k+\tau)|^2$ to the integral $\int_{t_0}^{t_0+T}dt A(t)$ where $A(t)$ designs the expansion \eqref{exp} of $|D(t,t+\tau)|^2$:
\bea\label{exp1}
\sum_{k=0}^{k=\nu-1}|D(t_k,t_k+\tau)|^2 = \omega_0^2\tau \int_{t_0}^{t_0+T}dt \zeta^2(t) &+& \omega_0^2\tau^3 \left\{\frac{1}{4}\int_{t_0}^{t_0+T}dt[\dot{\zeta}^2(t) +\omega_0^2\zeta^2(t)] -\frac{1}{3}[\zeta(t_0+T)\dot{\zeta}(t_0+T) -\zeta(t_0)\dot{\zeta}(t_0)]\right\} \nn\\
&+&{\cal O}(\tau^4).
\eea

We also expand \eqref{varpappendix} to the third order in $\tau$:
\be\label{exp2}
\bra \Delta^2 P(\phi_{t_{k+1},t_k} -\omega_0\tau)[t_{k+1},t_k]\ket = \frac{1}{4}\left(1- \tau^2{\cal K}^2 \right)\left(E+\sqrt{E^2-\frac{1}{4}}\right)^{-1} + \tau^2{\cal N} + {\cal O}(\tau^4),
\ee
 remembering that this is valid if $\tau$ is smaller than $\Omega_p^{-1}$, $p\geq2$ and $|\chi_q|^{-1}$, $q\geq 1$ and ${\cal K}^2:=\sum_n|K_n|^2$, ${\cal N}:=\sum_n|K_n|^2(N_n+1/2)$ (see Appendix \ref{considtimescale}).

Substituting the expressions \eqref{exp1} and \eqref{exp2} in the total quantum Fisher information we have
\bea
{\cal F}_Q^{Seq}(T,\tau) &=& \sum_{k=0}^{\nu-1} {\cal F}_Q(t_{k+1},t_k)\nn\\
&=& \sum_{k=0}^{\nu -1}\frac{|D(t_{k+1},t_k)|^2}{\bra \Delta^2 P(\phi^D_{t_{k+1},t_k} -\omega_0\tau)[t_{k+1},t_k]\ket}\nn\\
& =&\omega_0^2\frac{\xi(T,t_0) \tau + {\cal C}(T,t_0)\tau^3 +{\cal O}(\tau^4)}{\frac{{\cal E}^{-1}}{4} + \tau^2\left({\cal N}-\frac{{\cal E}^{-1}}{4}{\cal K}^2\right) + {\cal O}(\tau^4)},\label{fseqexp}
\eea
where ${\cal E}:=\left(E+\sqrt{E^2-\frac{1}{4}}\right)$, $\xi(T,t_0):=\int_{t_0}^{t_0+T}dt\zeta^2(t)$, and ${\cal C}(T,t_0):= \frac{1}{4}\int_{t_0}^{t_0+T}dt[\dot{\zeta}^2(t) +\omega_0^2\zeta^2(t)] -\frac{1}{3}[\zeta(t_0+T)\dot{\zeta}(t_0+T) -\zeta(t_0)\dot{\zeta}(t_0)]$ which simplifies to ${\cal C}(T,t_0):= \frac{1}{4}\int_{t_0}^{t_0+T}dt[\dot{\zeta}^2(t) +\omega_0^2\zeta^2(t)]$ if the force begins and ends  at $t_0$ and $t_f$ respectively. \\

From the expansion \eqref{fseqexp} one can easily find the optimal time interval $\tau_{\mathrm{opt}}$:
\bea
\tau_{\mathrm{opt}}=& \frac{1}{2\sqrt{3{\cal N}}}{\cal E}^{-1/2} &+ \frac{{\cal C}(T,t_0)+\xi{\cal K}^2}{16\sqrt{3}{\cal N}^{3/2}\xi(T,t_0)}{\cal E}^{-3/2} + {\cal O}({\cal E}^{-5/2}).\nn
\eea

 This result confirms the announced correlation between $\tau_{\mathrm{opt}}$ going to zero and $E$ going to infinity. Interestingly the leading term does not depend on the total available sensing window $[t_0;t_0+T]$ neither on the force time modulation $\zeta(t)$. For Markovian bath the dependence of $\tau_{\mathrm{opt}}$ is in ${\cal E}^{-1/3}$ \cite{pra}.\\

The corresponding total quantum Fisher information is
\bea
&&{\cal F}_Q^{\mathrm{Seq}}(T,\tau_{opt})= \frac{\sqrt{3}\xi(T,t_0)}{2\sqrt{\cal N}}{\cal E}^{1/2} + \frac{\sqrt{3}}{32{\cal N}^{3/2}}[2{\cal K}^2\xi(T,t_0) + 7/3{\cal C}(T,t_0)]{\cal E}^{-1/2} + {\cal O}({\cal E}^{-3/2}).\nn
\eea

If the effective time window during which the force is not null is $[t_i,t_f] \in [t_0,t]$, then we have $\xi(T,t_0)=\int_{t_i}^{t_f}dt\zeta^2(t) =\xi(t_f-t_i,t_i) $ and ${\cal C}(T,t_0):= \frac{1}{4}\int_{t_i}^{t_f}dt[\dot{\zeta}^2(t) +\omega_0^2\zeta^2(t)] = {\cal C}(t_f-t_i,t_i)$ and the total quantum Fisher information is just equal to ${\cal F}_Q^{Seq}(t_f-t_i,t_i)$: it is not prejudicial to starts and stops the sensing beyond the real time window of the force application. The exact knowledge of $t_i$ and $t_f$ is not necessary for sequential preparation-and-measurement scenario.\\

As a matter of comparison, we give the asymptotical behaviour of the optimal time interval and of the corresponding QFI when a term of 1st order in $\tau$ appears at the denominator of \eqref{qfipvar}. This happens when the time interval $\tau$ between each measurement is bigger than the evolution time scales of the bath correlation function (see Appendix \ref{nmvsm}) or when the dynamics is Markovian (which implies obviously that $\tau$ is bigger than the evolution time scales of the bath correlation function). In such situations Eq. \eqref{fseqexp} becomes
  \bea
{\cal F}_Q^{Seq}(T,\tau) & =&\omega_0^2\frac{\xi(T,t_0) \tau + {\cal C}(T,t_0)\tau^3 +{\cal O}(\tau^4)}{\frac{{\cal E}^{-1}}{4} + A\tau + {\cal O}(\tau^2)},
\eea
where $A$ is a coefficient appearing in situations described above ($A=\gamma(n_T+1/2)$ for Markovian dynamics), yielding 
\be
\tau_{opt} = \frac{1}{8A}{\cal E}^{-1/2}
\ee
and a bounded QFI,
\be
{\cal F}_Q^{\mathrm{Seq}}(T,\tau_{opt}) = \frac{\xi(T,t_0)}{3A} + {\cal O}({\cal E}^{-1}),
\ee
equivalent to the result in \cite{pra} for Markovian dynamics.

\section{Correlation function and time of the bath}\label{bathcorrelation}
The bath correlation function can be defined by the following expression: 
\be
C(t,t_0|t',t_0):=\frac{1}{2}{\rm Tr}_{SB}\{\rho_{SB}^0[B(t,t_0,F)B^{\dag}(t',t_0,F)+B^{\dag}(t',t_0,F)B(t,t_0,F)]\} - {\rm Tr}_{SB}[\rho_{SB}^0B(t,t_0,F)]{\rm Tr}_{SB}[\rho_{SB}^0B^{\dag}(t',t_0,F)]
\ee
where $B(t,t_0,F):=U^{\dag}(t,t_0,F)BU(t,t_0,F)$.\\

One can show the useful expression for $B(t,t_0,F)$:
\be
B(t,t_0,F) = B_0(t) -a_0(t)\dot{G}(t,t_0) + \int_{t_0}^t ds \dot{G}(t,s)B_0(s)e^{-i\omega_0(t-s)} +i\frac{F}{\sqrt{2}}\sum_n K_n e^{-i\omega_n(t-t_0)}D_n(t,t_0),
\ee
where $\dot{G}(t,s):= \frac{d}{dt}G(t,s)$, and $D_n(t,t_0)$ is defined in Appendix \ref{evolution}. \\

One gets for the bath correlation function, assuming that $\forall n$, ${\rm Tr}_B[\rho_B^0b_n]={\rm Tr}_B[\rho_B^0b_n^{\dag}]=0$, 
\be\label{bathcofct}
C(t,t_0|t',t_0) = C_{b}(t,t_0|t',t_0)+C_{I}(t,t_0|t',t_0),
\ee
where the first part 
\be
C_{b}(t,t_0|t',t_0)=e^{-i\omega_0(t-t')}C^0(t-t'),
\ee
corresponds to the bare correlation function of the bath without interaction with the probe, corresponding also to the Born approximation, and the second part,
\bea
C_I(t,t_0|t',t_0)&=& e^{-i\omega_0(t-t')}\Bigg\{\int_{t_0}^{t'}dsC^0(t-s)\dot{G}^{*}(t',s) +\int_{t_0}^tdsC^0(s-t')\dot{G}(t,s)\nn\\
&+&\left[\frac{1}{2}{\rm Tr}_S[\rho^0_S(a^{\dag}a+aa^{\dag})]-|{\rm Tr}_S(\rho_s^0a)|^2\right]\dot{G}(t,t_0)\dot{G}^{*}(t',t_0)
+ \int_{t_0}^tds\int_{t_0}^{t'}ds' C^0(s-s')\dot{G}(t,s)\dot{G}^{*}(t',s')\Bigg\},\nn\\
\eea
 gathers second and higher order terms coming from the interaction with the probe, involving the derivative of the response function of the bath $\dot{G}(t,s)$. The function $C^0(t-t')$ is defined above in Appendix \ref{considtimescale}, and $N_n={\rm Tr}_B[\rho_B^0b_n^{\dag}b_n]$. \\

Note that if one looks at the bath correlations at the beginning of the interaction between the bath and the probe, the correlation function is reduced to ($t'\rightarrow t_0$ in \eqref{bathcofct})
\bea
C(t,t_0|t_0,t_0) &=& e^{-i\omega_0(t-t_0)}\left[C^0(t-t_0) +\int_{t_0}^tdsC^0(s-t_0)\dot{G}(t,s)\right].
\eea

As detailed in Appendix \ref{considtimescale} the evolution time scale of $G(t,t_0)$ and $C^0(t-t')$ are of the order of $\Omega_p^{-1}$, $p\geq2$ and $|\chi_q|^{-1}$, $q\geq1$, and since \eqref{bathcofct} depends only on these two functions, the evolution time scale of the bath correlation function is also of the order of $\Omega_p^{-1}$ and $|\chi_q|^{-1}$. This is an important conclusion since it shows that the short time effects considered in this work are within the bath correlation time, justifying their classification as non-Markovian effects.\\

Note finally that under traditional Markovian approximation, including broad band limit, rotating wave approximation and Born approximation, the bath correlation function $C(t,t_0|t',t_0)$ becomes a delta Dirac function $\delta(t-t')$, implying that the correlation time is zero, and the impossibility of performing any measurement within this time.

\end{widetext}


\begin{thebibliography}{1} 
\bibitem{natphot} J. Aasi et al., Nature Photonics {\bf 7}, 613–619 (2013).
\bibitem{ligo} B. P. Abbott et al. (LIGO Scientific Collaboration and Virgo Collaboration) Phys. Rev. Lett. {\bf 116}, 061102 (2016).
\bibitem{impulsiveforce} D. Vitali, S. Mancini, and P. Tombesi, Phys. Rev. A {\bf 64}, 051401 (2001).
\bibitem{microopto} A. Pontin, M. Bonaldi, A. Borrielli, F. S. Cataliotti, F. Marino, G. A. Prodi, E. Serra, and F. Marin, Phys. Rev. A {\bf 89}, 023848 (2014).
\bibitem{aspel} M. R. Vaner et al., Proc. Natl. Acad. Sci. USA {\bf 108}, 16182 (2011).
\bibitem{biercuck} M. J. Biercuk, H. Uys, J. W. Britton, A. P. VanDevender, and J. J. Bollinger, Nat. Nanotechnol. {\bf 5}, 646 (2010).
\bibitem{ieee} K. S. Karvinen, M. G. Ruppert, K. Mahata and S. O. R. Moheimani, IEEE Trans. Nanotechnol. {\bf 13}, 1257-1265 (2014). 
\bibitem{nanotube} J. Moser, J. Guttinger, A. Eichler, M. J. Esplandiu, D. E. Liu, M. I. Dykman, and A. Bachtold, Nat. Nanotechnol. {\bf 8}, 493-496 (2013).
\bibitem{singlespin} D. Rugar, R. Budakian, J.H. Mamin, and B. W. Chui, Nature {\bf 430}, 329-332 (2004).
\bibitem{pt} A. C. Bleszynski-Jayich, W. E. Shanks, B. Peaudecerf, E. Ginossar, F. von Oppen, L. Glaz- man, and J. G. E. Harris, Science {\bf 326}, 272-275 (2009).
\bibitem{ca} U. Mohideen and A. Roy, Phys. Rev. Lett. {\bf 81}, 4549 (1998).
\bibitem{planck} I. Pikovski, M. R. Vanner, M. Aspelmeyer, M. S. Kim and C. Brukner, Nat. Phys. {\bf 8}, 393-397 (2012).
\bibitem{nori} M. Antognozzi, S. Simpson, R. Harnima, J. Senior, R. 
Hayward, H. Hoerber, M. R. Dennis, A. Y. Bekshaev, K. Y. Bliokh, and F. Nori, Nat. Phys. {\bf 12}, 731-735 (2016).
\bibitem{biology} M. A. Taylor and W. P. Bowen, Phys. Rep. {\bf 615}, 1-59 (2016).
\bibitem{euroj} Y. Gao, H. Lee, and Y. Lei Jia, Eur. Phys. J. D {\bf 68}:321 (2014).
\bibitem{plenio} A. W. Chin, S. F. Huelga, and M. B. Plenio, Phys. Rev. Lett. {\bf 109}, 233601 (2012).
\bibitem{matsuzaki} Y. Matsuzaki, S. C. Benjamin, and J. Fitzsimons, Phys. Rev. A {\bf 84}, 012103 (2011).
\bibitem{dd} A. Smirne, J. Kolodynski, S. F. Huelga, and R. Demkowicz-Dobrza\'nski, Phys. Rev. Lett. {\bf 116}, 120801 (2016).
\bibitem{pechukas} P. Pechukas, Phys. Rev. Lett. {\bf 73},1060 (1994); Phys. Rev. Lett. {\bf 75}, 3021 (1995).
\bibitem{fisher} R. A. Fisher, Math. Proc. Cambridge Philos. Soc. {\bf 22}, 700 (1925).
\bibitem{braunstein} S. L. Braunstein, J. Phys. A {\bf 25}, 3813 (1992).
\bibitem{cramer} H. Cramer,  {\it Mathematical Methods of Statistics} (Princeton University, Princeton, 1946).
\bibitem{rao} C. R. Rao, {\it Linear Statistical Inference and its Applications}, 2nd edn. (Wiley, New York, 1973).
\bibitem{tsang} M. Tsang, arXiv:1605.03799 (2016).
\bibitem{bcm96} S. L. Braunstein, C. M. Caves, and G. J. Milburn, Ann. Phys. {\bf 247}, 135 (1996).
\bibitem{paz} B. L. Hu, J. P. Paz, and Y. Zhang, Phys. Rev. D {\bf 45}, 2843 (1992).
\bibitem{pra} C. L. Latune, B. M. Escher, R. L. de Matos Filho, and L. Davidovich, Phys. Rev. A {\bf 88}, 042112 (2013).
\bibitem{caves} S. L. Braunstein and C. M. Caves, Phys. Rev. Lett. {\bf 72} 3439 (1994).
\bibitem{paris} M. G. A. Paris, Int. J. Quant. Inf. {\bf 7}, 125 (2009).
\bibitem{bruno} B. M. Escher, R. L. de Matos Filho and L. Davidovich, Nat. Phys. {\bf 7}, 406 (2011).
\bibitem{nicim} B. M. Escher, L. Davidovich, N. Zagury, and R. L. de Matos Filho, Phys. Rev. Let. {\bf 109}, 190404 (2012).
\bibitem{scutaru} H. Scutaru, Journal of Physics A {\bf 31}, 3659 (1998).
\bibitem{pinel} O. Pinel, P. Jian, N. Treps, C. Fabre, and D. Braun, Phys. Rev. A {\bf 88}, 040102 (2013).
\bibitem{monras} A. Monras, arXiv:1303.3682 (2013).
\bibitem{jiang} Zhang Jiang, Phys. Rev. A {\bf 89}, 032128 (2014).
\bibitem{gardiner} C. W. Gardiner and P. Zoller, {\it Quantum Noise}, (Second Enlarged Edition, Springer - Velag Berlin Heidelberg New-York, 2000). 
\bibitem{screport} Y.-­R. Zhang and H. Fan, Sci. Rep. {\bf 5}:11509 (2015).
\bibitem{zeno} A. H. Kiilerich and K. Molmer, Phys. Rev. A {\bf 92}, 032124 (2015).
\bibitem{tsang1} M. Tsang, H. M. Wiseman, and C. M. Caves, Phys. Rev. Lett. {\bf 106}, 090401 (2011).
\bibitem{tsang2} M. Tsang, New J. Phys. {\bf 15}, 073005 (2013).
%\bibitem{brca} S. L. Braunstein, C. M. Caves, Phys. Rev. Lett. {\bf 72}, 3439 (1994).
\end{thebibliography}
\end{document}